\documentclass[reprint,aps,amsmath,amssymb,nofootinbib]{revtex4-2}

\usepackage{bbold}
\usepackage{tabu}
\usepackage{graphicx,subcaption}
\usepackage{amsfonts}
\usepackage{tensor}
\usepackage{yhmath}
\usepackage{epsfig}
\usepackage{slashed}
\usepackage{pdfpages}
\usepackage{psfrag}
\usepackage{graphicx,epstopdf}
\usepackage{soul}

\colorlet{darkblue}{blue!70!black}
\usepackage[colorlinks=true,
urlcolor=darkblue,
linktocpage=true,
linkcolor=darkblue,
citecolor=green]{hyperref}
\usepackage{array}
\usepackage{mathtools}
\usepackage{braket}
\usepackage{enumitem}
\usepackage{bm}
\usepackage[thinlines]{easytable}
\usepackage{cancel}
\usepackage{leftidx}
\usepackage{caption}
\usepackage{tikz}
\usetikzlibrary{patterns}
\usetikzlibrary{decorations.markings}
\usetikzlibrary{decorations.pathmorphing}
\usetikzlibrary{decorations.pathreplacing}
\usetikzlibrary{shapes.misc} 
\usetikzlibrary{shapes.geometric}
\usetikzlibrary {arrows.meta,bending,positioning}
\tikzset{dot/.style = {circle, fill, minimum size=#1,
              inner sep=0pt, outer sep=0pt}}
\tikzset{cross/.style={cross out, draw, 
         minimum size=2*(#1-\pgflinewidth), 
         inner sep=0pt, outer sep=0pt}}

\newcommand{\mC}{\mathcal{C}}

\newcommand{\mO}{\mathcal{O}}
\newcommand{\mL}{\mathcal{L}}

\newcommand{\mD}{\mathcal{D}}
\newcommand{\mN}{\mathcal{N}}
\newcommand{\mB}{\mathcal{B}}

\newcommand{\eps}{\epsilon}
\newcommand{\vareps}{\varepsilon}

\newcommand{\no}{\nonumber}
\newcommand{\der}{\partial}
\newcommand{\be}{\begin{equation}}
\newcommand{\ee}{\end{equation}}

\graphicspath{{./figures/}}

\usepackage{geometry}
\newgeometry{includefoot,left=2cm,right=2cm,bottom=1cm,top=2cm}

\makeatletter
\AtBeginDocument{\let\LS@rot\@undefined}
\makeatother

\begin{document}

\title{Non-Factorizing Interface in the Two-Dimensional Long-Range Ising Model}

\author{Dongsheng Ge}\email{dge@het.phys.sci.osaka-u.ac.jp}
\affiliation{
Department of Physics, The University of Osaka,\\
 Machikaneyama-Cho 1-1, Toyonaka 560-0043, Japan}

\author{Yu Nakayama}\email{yu.nakayama@yukawa.kyoto-u.ac.jp}
\affiliation{
 Yukawa Institute for Theoretical Physics, Kyoto University,\\
Kitashirakawa Oiwakecho, Sakyo-ku, Kyoto 606-8502, Japan}

\date{\today}


\begin{abstract}
The factorization proposal claims that the co-dimension one ``pinning defect", on which a local relevant operator is integrated, factorizes the space into two halves in general conformal field theories in the infrared limit. In this letter, we study a two-dimensional long-range Ising model at criticality with a line defect or an interface, which physically corresponds to changing the local temperature on it. We show that in the perturbative regime, it is not factorizing even in the infrared limit. An intuitive explanation of the non-factorization is that the long-range Ising model is equivalent to a local conformal field theory in higher dimensions. In this picture, the space is still connected through the ``extra dimension" across the defect line. 

\end{abstract}


\maketitle


\emph{1. Introduction.--}
What happens if a heater (or cooler) is applied on a co-dimension one hyperplane in the critical Ising model? Will the criticality remain on this interface, or will it be destroyed? In the two-dimensional critical short-range Ising model, changing the local temperature on the line is marginal; Affleck and Oshikawa gave an exact description, where the defect line remains critical for a whole range of temperatures \cite{Oshikawa:1996ww,Oshikawa:1996dj}. Furthermore, they showed that a localized renormalization group (RG) flow triggered by a relevant deformation from the local magnetic field makes the entire space factorize into two halves in the infrared (IR), with no energy exchange between them \cite{Quella:2006de}.

In the three-dimensional Ising model, changing the temperature or applying the magnetic field locally on the interface is relevant,  and we expect that the interface will factorize the space into halves in the IR. Since so far the bulk theory can be studied only through numerical analysis such as the Monte-Carlo simulations \cite{Hasenbusch:2021tei}, conformal bootstrap \cite{El-Showk:2012cjh,El-Showk:2014dwa,Nakayama:2016cim,Kos:2016ysd,Chang:2024whx}, or fuzzy sphere \cite{Zhu:2022gjc,Hu:2023xak}, the explicit demonstration may be difficult.\footnote{There has been interesting works studying the boundary three-dimensional Ising model from the conformal bootstrap perspective appeared in \cite{Gliozzi:2015qsa}, and recent works using the fuzzy sphere can be found in \cite{Zhou:2024dbt,Dedushenko:2024nwi}. } 
Nonetheless, in the closely related $O(N)$ model with a planar interface, some evidence of exhibiting such a  factorization  has been reported in \cite{Krishnan:2023cff,Toldin:2024pqi}. 

More recently, Popov and Wang proposed that the factorization property should hold for a conformal field theory (CFT) in general dimensions with a co-dimension one ``pinning defect'' \cite{Popov:2025cha}.\footnote{A partial list of ``pinning'' type models can be found in \cite{Parisen_Toldin_2017,Cuomo:2021kfm,Rodriguez-Gomez:2022gbz,Gimenez-Grau:2022czc,Giombi:2022vnz,Barrat:2023ivo,Nishioka:2022qmj,Pannell:2023pwz,Pannell:2024hbu,Zhou:2023fqu,SoderbergRousu:2023zyj,Diatlyk:2024zkk,Gabai:2025zcs,Shachar:2022fqk,Trepanier:2023tvb,Giombi:2023dqs,Raviv-Moshe:2023yvq,Diatlyk:2024ngd,deSabbata:2024xwn,Brax:2023goj,Harribey:2023xyv,Harribey:2024gjn,Giombi:2024zrt,Bianchi:2024eqm}.} They gave a formal proof and claimed that an RG flow triggered by the pinning field deformation shall lead the interface to factorize the space in the IR limit.

These developments motivate us to ask if the factorization is an intrinsic property for a pinning-type interface CFT \cite{Wong:1994np,Oshikawa:1996ww,Oshikawa:1996dj,Cardy:1989ir,McAvity:1995zd,Bachas:2001vj} in the IR limit. 
We pay attention to a special type of CFT in two dimensions, the long-range Ising (LRI) model, which is non-local and does not possess the Virasoro symmetry.\footnote{The absence of the Virasoro symmetry is due to the lack of the local stress tensor: For this reason, it is sometimes called conformal theory instead of conformal {\it field} theory. The proof of the Virasoro symmetry by Polchinski \cite{Polchinski:1987dy,Nakayama:2013is,Nakayama:2019xzz} does not apply due to the absence of the local stress tensor.}
Surprisingly, we will show that changing the local temperature on the interface does not factorize the space. An intuitive explanation of the non-factorization is that the LRI model is equivalent to a local CFT in higher dimensions. In this picture, the space is still connected through the ``extra dimension" across the defect line.

The primary focus of our letter is to study the Landau-Ginzburg (LG) description of the LRI model with the introduction of  a localized massive deformation on the co-dimension one defect:
\be\label{eq:actionLRI}
S = \frac{\mN_s}{2} \int d^ d y\, \phi\, \mL_s \phi + \frac{h_0}{4!} \int d^d  y\, \phi^4
+  \frac{g_0}{2} \int d^r  z \, \phi^2
\,, 
\ee
where the kinetic term is the  ``fractional Laplacian'' $ \mL_s = (-\der^2)^{s/2}$. We assume $s$ is slightly above $d/2$, $s = d/2+\vareps~(0<\vareps\ll1)$, such that the interaction $\phi^4$ is weakly relevant and  can be studied perturbatively in $\vareps$.\footnote{The model without the line defect is the LRI model, for instance, reviewed in \cite{Paulos:2015jfa} and earlier works \cite{Fisher:1972zz,Sak:1973oqx}. For $d/2<s <s^*$, the critical theory is non-trivial and non-Gaussian, which is supposed to be in the same universality class as the critical point of the LRI lattice model. We are interested in the region where  $s$ is slightly above $d/2$, while an up-to-date discussion on the transition from the LRI model to the short-range Ising (SRI) model where $s$ approaches $s^*$ from below appeared in \cite{Behan:2017dwr,Behan:2017emf}.} 
The classical dimension of the LG field is $\Delta_\phi = \frac{d-s}{2}$.
Eventually, we will be interested only in the co-dimension one case $r=d-1$. For the deformation by $\phi^2$ on the defect to be relevant $r>d-s$, this only leaves us with the choice $r=d-1=1$ or $r=d-1=2$ given $d\,,r\in \mathbb{Z}^+$ and $d<4$. In this letter, we focus on the former choice so that the coupling $h_0$ and $g_0$ can be perturbative simultaneously.\footnote{We thank {\it Tatsuma Nishioka} for discussion on the latter possibility.} 
This choice of $d=2,\, r=1$  physically corresponds to the two-dimensional LRI model with a line defect. Note that within physical dimensions, we can still tune $\varepsilon$ so that perturbative computation is valid.

While the LRI model has a non-local kinetic term, the so-called Caffarelli-Silvestre (CS) trick \cite{Caffarelli_2007} can be applied to map the same system to a {\it local} conformal field theory in $(2-s)$-higher dimensions with (multiple) defects.  With this trick, the LG field $\phi(y)$ is lifted to the  field $\Phi(x_\perp,y)$, where $\phi(y)$ can be viewed as the Dirichlet boundary condition $\phi(y) = \Phi(x_\perp=0,y)$.
The equivalent description is  
\begin{align}\label{eq:actionCD}
S'= \frac{1}{2} \int d^D x\, (\der_M \Phi)^2 + \frac{h_0}{4!} \int_{x_{\perp} = 0} d^d  y\, \Phi^4
\no\\
+ \frac{g_0}{2} \int_{x_{\perp}=0\,,y_{\perp} = 0} d^r  z\, \Phi^2
\,,
\end{align}
where $D = d+2-s = \frac{d}{2} +2 -\vareps$. 
The coordinates here are denoted as $x=(  x_\perp,y)$ where $x_\perp$ are the ``extra'' dimensions to the original $d$-dimensional LRI theory, and $y=(y_\perp,z)$, where $y_\perp$ are the transverse directions to the defect in the LRI model.\footnote{This model, in the end, coincides with the composite defect model studied in \cite{Ge:2024hei} by one of the authors.} In this letter, we use this equivalent formulation to employ the local field theory techniques.

\begin{figure}
    \begin{tikzpicture}[baseline=(vert_cent.base)]
       \node (vert_cent) at (0,1) {$\phantom{\cdot}$};
        \node[yshift=6pt] at (0,1.25) {$x$};
        \draw[draw=none, fill = gray!40, opacity=0.7] (-.6,0.3) -- (1, 0.3) -- (0.6, -0.3) -- (-1,-0.3) -- (-.6,0.3);
        \node [fill, shape=rectangle, minimum width=4pt, minimum height=4pt, inner sep=0pt, anchor=center] at (0,0) {};
        \draw[dashed] (0,0) edge[bend left=40] (0,1.25) ;
        \draw[dashed] (0,0) edge[bend left=20] (0,1.25) ;
        \draw[dashed] (0,0) edge[bend right=20] (0,1.25) ;
        \draw[dashed] (0,0) edge[bend right=40](0,1.25) ;
    \end{tikzpicture}
    \begin{tikzpicture}[baseline=(vert_cent.base)]
     \node (vert_cent) at (0,1) {$\phantom{\cdot}$};
     \node[yshift=6pt] at (0,0.6) {$+$};
    \end{tikzpicture}
      \begin{tikzpicture}[baseline=(vert_cent.base)]
        \node (vert_cent) at (0,1) {$\phantom{\cdot}$};
        \node[yshift=6pt] at (0,1.25) {$x$};
        \draw[draw=none, fill = gray!40, opacity=0.7] (-.6,0.3) -- (1, 0.3) -- (0.6, -0.3) -- (-1,-0.3) -- (-.6,0.3);
        \node [fill, shape=rectangle, minimum width=4pt, minimum height=4pt, inner sep=0pt, anchor=center] at (-.5,0) {};
        \node [fill, shape=rectangle, minimum width=4pt, minimum height=4pt, inner sep=0pt, anchor=center] at (.5,0) {};
        \draw[dashed] (-.5,0) edge[bend left=40] (0,1.25) ;
        \draw[dashed] (-.5,0) edge[bend left=20] (0,1.25) ;
        \draw[dashed] (.5,0) edge[bend right=20] (0,1.25) ;
        \draw[dashed] (.5,0) edge[bend right=40](0,1.25) ;
        \draw[dashed] (.5,0) edge[bend right=20] (-.5,0) ;
        \draw[dashed] (.5,0) edge[bend right=40](-.5,0) ;
    \end{tikzpicture}
\caption{Diagrams contributing to the one-loop order of the one-point function $\langle \Phi^4(x) \rangle$, with the solid square being the vertex of the quartic coupling $h_0$. }
\label{fig:diagrams1}
\end{figure}
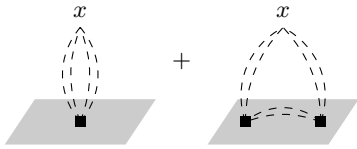

\emph{2. Long-range Ising model.--}
To start with, we turn off the localized massive deformation $g_0$ for the moment and study the bulk theory.
Such a theory can be renormalized
by keeping the following one-point function finite as $\vareps \to 0$ 
\be
\langle [\Phi^4](x) \rangle = \text{finite}\,,\footnote{For the parent field $\Phi$, as the wavefunction renormalization is trivial, we do not distinguish the bare field and the renormalized one. }
\ee
which involves two diagrams up to the second order in the coupling $h_0$, as in fig. \ref{fig:diagrams1}.  In a free bulk, the wavefunction renormalization is trivial, such that the bulk operator $\Phi^4 = [\Phi^4]$ up to normal ordering.
From now on, normal ordering of the composite operators is always assumed: we do not include any self-contractions of $\Phi$ in perturbative expansions.
The renormalization condition, for which we use minimal subtraction (MS) scheme, 
gives rise to the RG beta function for the coupling $h$:
\begin{align}\label{eq:betah}
   \beta_h = -2 \vareps h 
   +\frac{3  \Gamma \left(\frac{d}{4}\right)^2}{16 \pi ^2 \Gamma
   \left(\frac{d}{2}\right)}h^2 +\cdots
  \,,
\end{align}
which allows a non-trivial IR fixed point,
\begin{align}\label{eq:FPh}
h_* = \frac{32 \pi ^2 \Gamma \left(\frac{d}{2}\right)}{3  \Gamma \left(\frac{d}{4}\right)^2} \vareps + O(\vareps^2)\,.
\end{align}
At this fixed point, the theory describes the critical $d$-dimensional LRI model, which is conformally invariant \cite{Paulos:2015jfa}.
The LG field $\phi$ and composite operators $\phi^n$  living on the plane $x_\perp=0$ are  subject to the Dirichlet boundary conditions
$
\phi^n ( y) = \Phi^n ( y,x_\perp =0)
$.
They acquire an anomalous dimension at the LRI fixed point, which can be obtained in terms of their 
multiplicative renormalization factor $Z_{\phi^n}$, defined as $\phi^n \equiv Z_{\phi^n} [\phi^n]$ with $[\phi^n]$ being the renormalized composite operators. Perturbatively the renormalization factor can be computed as
\be
Z_{\phi^n} = 1 - \frac{ (n-1) n \Gamma \left(\frac{d}{4}\right)^2}{64 \pi ^2 \Gamma
   \left(\frac{d}{2}\right)} \frac{h}{\vareps} +\dots\,,
\ee
which gives the anomalous dimension at the fixed point 
\begin{align}
\gamma_{\phi^n}^* &=\frac{\der\ln Z_{\phi^n}}{\der\ln \mu} \Big|_{h = h_*} 
= \frac{ (n-1) n \Gamma \left(\frac{d}{4}\right)^2}{32 \pi ^2 \Gamma
   \left(\frac{d}{2}\right)} h_* +\dots\no\\
   &= \frac{(n-1)n}{3}\vareps + O(\vareps^2)\,.
\end{align}
Among those composite operators, there are two protected ones, $[\phi]$ and $[\phi^3]$ \cite{Paulos:2015jfa}. At this fixed point,
(i) $\gamma_\phi^*=0$ vanishes to all-loop order, due to the non-locality of its kinetic term, such that wavefunction renormalization is simply the identity (proved in \cite{Lohmann:2017qyq});
(ii) $\gamma_{\phi^3}^* = 2 \vareps$ to all-loop order, this is a result of the e.o.m. 
from the $D$-dimensional local field theory point of view
\be
\Box \Phi (x)= \frac{h}{3!} \delta^{(D-d)}(x_\perp) [\phi^3](y)~~~\Rightarrow~~~\Delta_{\phi^3} = \frac{d}{2}-\frac{s}{2}\,.
\ee
These two operators $[\phi]$ and $[\phi^3]$ form a shadow pair in the $d$-dimensional LRI theory \cite{Behan:2017dwr},
\be
\Delta_{\phi^3}  + \Delta_\phi  = 2 \Delta_\phi +s =d\,.
\ee
 

\emph{3. Non-local Gaussian CFT.--}
It is also interesting to consider the localized massive deformation alone by turning off the 
quartic coupling $h_0$. In this case, the model \eqref{eq:actionCD} becomes a $D$-dimensional free scalar theory with a localized mass term and  $d$ becomes fictitious. This is a Gaussian theory that can be solved completely. 
Using a similar renormalization procedure by requiring 
$
\langle [\Phi ^2] (x) \rangle = \text{finite}
$
as $\vareps \to 0$,
\begin{figure}
    \begin{tikzpicture}[baseline=(vert_cent.base)]
       \node (vert_cent) at (0,1) {$\phantom{\cdot}$};
        \node[yshift=6pt] at (0,1.) {$x$};
        \draw[draw=none, fill = gray!40, opacity=0.7] (-.4,0.3) -- (0.8, 0.3) -- (0.4, -0.3) -- (-0.8,-0.3) -- (-.4,0.3);
        \draw[thick] (-0.6,0)--(0.6,0);
        \filldraw (0,0) circle (2pt);
        \draw[dashed] (0,0) edge[bend left=20] (0,1.) ;
        \draw[dashed] (0,0) edge[bend right=20] (0,1.) ;
    \end{tikzpicture}
    \begin{tikzpicture}[baseline=(vert_cent.base)]
     \node (vert_cent) at (0,1) {$\phantom{\cdot}$};
     \node[yshift=6pt] at (0,0.4) {$+$};
    \end{tikzpicture}
      \begin{tikzpicture}[baseline=(vert_cent.base)]
        \node (vert_cent) at (0,1) {$\phantom{\cdot}$};
        \node[yshift=6pt] at (0,1.) {$x$};
        \draw[draw=none, fill = gray!40, opacity=0.7] (-.6,0.3) -- (1, 0.3) -- (0.6, -0.3) -- (-1,-0.3) -- (-.6,0.3);
        \draw[thick] (-0.8,0)--(0.8,0);
        \filldraw (-.5,0) circle (2pt);
        \filldraw (.5,0) circle (2pt);       
        \draw[dashed] (-.5,0) edge[bend left=20] (0,1.) ;
        \draw[dashed] (.5,0) edge[bend right=20] (0,1.) ;
        \draw[dashed] (.5,0) edge[bend right=40](-.5,0) ;
    \end{tikzpicture}
    \begin{tikzpicture}[baseline=(vert_cent.base)]
     \node (vert_cent) at (0,1) {$\phantom{\cdot}$};
     \node[yshift=6pt] at (0,0.4) {$+$};
    \end{tikzpicture}
 \begin{tikzpicture}[baseline=(vert_cent.base)]
        \node (vert_cent) at (0,1) {$\phantom{\cdot}$};
        \node[yshift=6pt] at (0,1.) {$x$};
        \draw[draw=none, fill = gray!40, opacity=0.7] (-.7,0.3) -- (1.1, 0.3) -- (0.7, -0.3) -- (-1.1,-0.3) -- (-.7,0.3);
        \draw[thick] (-0.9,0)--(0.9,0);
        \filldraw (-.7,0) circle (2pt);
        \filldraw (-.,0) circle (2pt);
        \filldraw (.7,0) circle (2pt);       
        \draw[dashed] (-.7,0) edge[bend left=20] (0,1.) ;
        \draw[dashed] (.7,0) edge[bend right=20] (0,1.) ;
        \draw[dashed] (.7,0) edge[bend right=40](-.,0) ;
        \draw[dashed] (-.7,0) edge[bend left=40](-.,0) ;
    \end{tikzpicture}
    \begin{tikzpicture}[baseline=(vert_cent.base)]
     \node (vert_cent) at (0,1) {$\phantom{\cdot}$};
     \node[yshift=6pt] at (0,0.4) {$+$};
     \node  at (0.5,0.6) {$\cdots$};
    \end{tikzpicture}
\caption{Loop-like diagrams contributing to the one-point function $\langle \Phi^2(x) \rangle$, with the solid dot representing the vertex of the quadratic coupling $g_0$. }
\label{fig:diagrams2}
\end{figure}
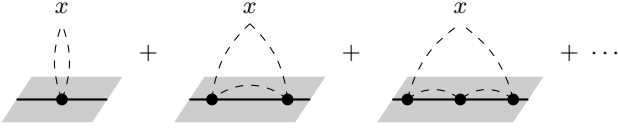
we sum up all the loop-like diagrams depicted in fig. \ref{fig:diagrams2}
\begin{align}\label{eq:Phi2regu}
\langle[ \Phi ^2] (x) \rangle 
= -g_0  &\int \frac{d^D p}{(2\pi)^{D}}\int \frac{d^D p'}{(2\pi)^{D-r}} \frac{e^{i(p-p')x}}{p^2}\no\\
 &\frac{1}{1+ g_0 \frac{a_{D,r}}{\bar p^{2+r-D}}}\frac{\delta^{(r)} (\bar p' - \bar p)}{p'^2}\,.
\end{align}
Note that since the bulk is a free theory, this bulk operator does not have any wavefunction renormalization, namely  $\Phi^2 = [\Phi^2]$ up to normal ordering.
The ``barred'' momenta are the parallel components in the $r$-dimensional space, and $ a_{D,r} \equiv  (4\pi)^{\frac{r-D}{2}}\Gamma\left( \frac{2+r-D}{2} \right)$. We focus on the case $\varepsilon = 2 -D+r >0$ so that the localized ``mass" deformation is relevant.
The  renormalization condition \eqref{eq:Phi2regu} leads to the exact beta function for the localized ``mass" deformation
\be
g_0 = \mu^\vareps \frac{g}{1-\frac{g}{2\pi \vareps}}
~~~\Rightarrow ~~~ 
\beta_g =  \frac{g (g-2\pi \vareps)}{2\pi}\,,
\ee
admitting an exact non-trivial IR fixed point at 
\be\label{eq:gGFP}
g_* =2\pi \vareps\,.
\ee
At this fixed point, the field $\hat \phi$, defined as the Dirichlet boundary condition of the field $\Phi$ on the $r$-dimensional hyperplane 
$
\hat \phi (z) = \Phi(z; x_\perp = 0, y_\perp = 0)
$,
acquires an anomalous dimension, though the kinetic term remains non-local.
Its anomalous dimension can be obtained by calculating the two-point function of the renormalized field $[\hat \phi]$. To all-loop order, this gives
\begin{align}\label{eq:massive2pt}
\langle [\hat\phi](z_1) [\hat \phi](z_2) \rangle 
&=  \frac{1}{Z_{\hat\phi}^2}\int \frac{d^{r} \bar p}{(2\pi)^r} \frac{e^{i z_{12} \bar p}}{ 
{ \bar p^{2+r-D} \over a_{D,r}} + g_0}\,.
\end{align}
This calculation can be done, first in the parallel momentum space using the propagator  
(I.4) and then Fourier transforms back to the position space.
The multiplicative renormalization factor to all-loop order is obtained as
\be
Z_{\hat \phi} = 1-\frac{g}{2\pi \vareps} = \frac{\mu^\vareps g }{g_0 }\,,
\ee
using the fixed point value, \eqref{eq:gGFP} gives the anomalous dimension of $\hat \phi$ 
\be\label{eq:anophi}
\gamma_{\hat \phi}^* =
\vareps \,.
\ee
This is consistent with the large-distance behavior\footnote{This integral can also be explicitly performed by using the Mittag-Leffler function \cite{BookML} (in particular when $r=1$).} of the all-loop correlator \eqref{eq:massive2pt}. By expanding it for small parallel momentum $\bar p \ll \mu$, 
\be
\langle [\hat\phi](z_1) [\hat \phi](z_2) \rangle  \stackrel{z_{12}\gg 1/\mu}{\sim} \frac{\Gamma(r/2)}{4\pi^{1+r/2}} |z_{12}|^{-r-\vareps}\,.
\ee
The anomalous dimension can be read immediately, giving the same value as \eqref{eq:anophi}. 
Alternatively, the anomalous dimension of $\hat \phi$ can be read directly from the e.o.m. $\Box\Phi(x) = g \,\delta^{(D-r)}(x_\perp;y_\perp) [\hat \phi](z)$, which gives \eqref{eq:anophi} as well, since the operator $\hat \phi$ is protected. Noting that the operators $\Phi$ and $\hat\phi$ satisfy a shadow relation in the $r$-dimensional theory at the IR fixed point,
\be
\Delta_{\Phi} + \Delta_{\hat \phi} = r\,.
\ee
One can also obtain the anomalous dimension for the composite operators $\hat \phi^n$
\begin{align}\label{eq:GHphiN}
Z_{\hat \phi^n}
 =\left( Z_{\hat \phi}\right)^n~~~ \Rightarrow~~~
 \gamma_{\hat \phi^n} ^*= n\,\vareps\,.
\end{align}
As the two-point function of $\hat\phi^n$ factorizes into the $n$-th power of the two-point function of $\hat \phi$, $\langle \hat\phi^n(z_1) \hat \phi^n(z_2) \rangle = \left( \langle \hat\phi(z_1) \hat \phi(z_2) \rangle \right) ^n$.

Let us briefly discuss the physical implications. Unlike 
the short-range Gaussian theory, adding localized mass term does not gap out the defect line, and a non-trivial critical behavior remains. In the fictitious $d$-dimensional non-local Gaussian theory, we can show that 
the interface (i.e. $d=2$ and $r=1$) is not factorizing, for instance setting $h = 0$ in 
sec. II of the supplemental materials \cite{supp}.
The essence of such non-factorization roots in the fractional kinetic term, as can be seen in a simpler quantum mechanical system illustrated in sec. III of the supplemental materials \cite{supp}.
This should be contrasted with the short-range Gaussian theory, where the massive deformation on the interface factorizes the space in the IR limit as expected from the factorization proposal.


\begin{figure}
    \begin{tikzpicture}[baseline=(vert_cent.base)]
       \node (vert_cent) at (0,1) {$\phantom{\cdot}$};	
        \node[yshift=6pt] at (0,1.) {$x$};
        \draw[draw=none, fill = gray!40, opacity=0.7] (-.6,0.3) -- (1, 0.3) -- (0.6, -0.3) -- (-1,-0.3) -- (-.6,0.3);
        \draw[thick] (-0.8,0)--(0.8,0);
  	 \node [fill, shape=rectangle, minimum width=4pt, minimum height=4pt, inner sep=0pt, anchor=center] at (-.5,0.15) {};
	 \filldraw (.5,0) circle (2pt); 
        \draw[dashed] (-0.5,0.15) edge[bend left=20] (0,1.) ;
        \draw[dashed] (-.5,0.15) edge[bend right=20] (0,1.) ;
        \draw[dashed] (-0.5,0.15) edge[bend left=20] (.5,0) ;
        \draw[dashed] (-.5,0.15) edge[bend right=20] (.5,0) ;
    \end{tikzpicture}
\caption{The mixed diagram contributing to  $\langle \Phi^2(x) \rangle$ at the second order in the couplings, with the square and dot representing the vertices of $h_0$ and  $g_0$ respectively. }
\label{fig:diagrams3}
\end{figure}
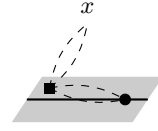

\emph{4. Long-range Ising with a line defect.--}
By turning on simultaneously the quartic coupling $h_0$ and the localized massive deformation $g_0$, we can now study if the co-dimension one defect will factorize the space in halves in the LRI model at the criticality.
While the beta function for $h$ remains unchanged as in \eqref{eq:betah}, the beta function for $g$ receives corrections from $h$. At the second order in the couplings, an extra mixed diagram (fig. \ref{fig:diagrams3})  contributes to the beta function of $g$ (taking $r=d-1=1$)
\begin{align}
\beta_g &=- \vareps g + \frac{gh +8g^2}{16\pi}\,.
\end{align}
Then the IR fixed point of $g$ is shifted by the fixed point value of $h$ in \eqref{eq:FPh} 
\be\label{eq:lineFP}
g_* =2 \pi  \vareps - \frac{h_*}{8}
   = \frac{2\pi \vareps}{3} + O(\vareps^2)\,.
\ee
The bulk composite operators $\phi^n$ remain the same as in the LRI model, but for those composite defect operators $\hat \phi^n$ living on the line, their anomalous dimensions get further corrected by the quartic coupling as compared to the non-local Gaussian case. 
At the new fixed point, 
\begin{align}\label{eq:ADhatphin}
 \gamma_{\hat \phi^n}^*=  \frac{ (n-1) n \Gamma \left(\frac{d}{4}\right)^2}{32 \pi ^2 \Gamma
   \left(\frac{d}{2}\right)} h_* + \frac{n\,g_*}{2\pi} = \frac{n^2}{3}\vareps+ O(\vareps^2)\,,
\end{align}
scales as $n^2$ at the leading order in $\vareps$. 

Given these composite operators on the line defect, we now study the bulk-defect two-point functions. This gives a perturbative check of the conformal invariance of the line defect, which will be detailed in the next section. 
Assuming the system is conformally invariant, these correlators should have the following form\footnote{Note here that the scaling operators $[\phi]^n$ are inserted away from the interface. A special case is $n=4$, where a true scaling operator is a mixture of $[\phi]^4(y)$ and $\delta(y_\perp)[\hat \phi^2](z)$ discussed in detail in sec. IV.A of the supplemental materials \cite{supp}.} 
\begin{align}\label{eq:bulkdefect}
\langle  [ \phi^n ](  y) [\hat \phi ^n](  z) \rangle  = \frac{\mN_{\phi^n} \mN_{\hat \phi^n}\mB_{\phi^n\hat \phi^n}}{|y_\perp|^{\gamma_{\phi^n} -\gamma_{\hat \phi^n}} |y -z |^{2\Delta_{\hat \phi^n}}} \,,
\end{align}
where  $\mN_{\phi^n}$ and  $\mN_{\hat \phi^n}$ are the two-point function normalizations for $\phi^n$ and $\hat \phi^n$ respectively,
\begin{align}
\langle [ \phi^n ] (  y_1) [ \phi^n] (  y_2) \rangle &= \frac{\mN_{\phi^n}^2}{|y_{12}|^{2(n \Delta_\phi +\gamma_{\phi^n})}}\,,\\
\langle [\hat \phi^n  ] (  z_1) [ \hat\phi^n] (  z_2) \rangle &= \frac{\mN_{\hat\phi^n}^2}{|z_{12}|^{2(n \Delta_\phi +\gamma_{\hat\phi^n})}}\,.
\end{align}
 Here the two-point function for $\phi^n$ is the one without the presence of the line defect. The additional factor $\mB_{\phi^n\hat \phi^n}$ encodes the bulk-to-defect OPE data. 
The explicit computation shows that to the first order corrections in $\vareps$, the bulk-defect two-point function exhibits the conformal form, with the related conformal data  given as
\begin{align}
&\mN_{\phi^n}^2 = \frac{n!}{\left( 4\pi \right)^{n}} \left(1  -\frac{ n (2 n-5)\vareps}{6} \left(\ln(4\pi)+\gamma
\right) \right)\,,\\
&\mN_{\hat\phi^n}^2 =\frac{n!}{\left( 4\pi \right)^{n}} \left(1  -\frac{ n (2 n-3)\vareps}{6} \left(\ln(4\pi)+\gamma\right) \right)\,,\\
\label{eq:Btdcoeff}
&\mB_{\phi^n\hat \phi^n} =1-  \frac{n\, \vareps}{3}   \ln 2\,,
\end{align}
where $\gamma$ is the Euler-Mascheroni constant. This provides the first non-trivial check of conformal invariance of the line defect, whose formal proof will be given in the following section.


\emph{5. Non-factorizing conformal interface.--}
The two-dimensional LRI with a line defect discussed above gives rise to a scale-invariant interface theory, as indicated by the existence of a non-trivial IR fixed point \eqref{eq:lineFP} with the line deformation.  In fact, such an interface theory is conformally invariant, \textit{albeit} only in the global sense due to its non-locality. The conformal invariance of the two-dimensional LRI model itself was proved in \cite{Paulos:2015jfa}. Adding an interface breaks part of its translational symmetries as well as special conformal symmetries. However, the residual conformal symmetries, as we shall show, are still preserved on the interface.
Starting directly from the LRI model would be hard, as a lack of a conserved local stress tensor \cite{Behan:2017dwr}. From the higher-dimensional theory perspective, however, the conformal invariance of the LRI model with the line defect becomes relatively straightforward.

We follow the procedures in \cite{Brown:1979pq,Paulos:2015jfa}. The idea is that the conformal Ward-Takahashi identities are broken along the RG flow, but should be restored at the fixed point due to the lack of the candidate for the ``virial current" that is necessary to realize scale invariance without conformal invariance. This suggests that the brokenness is likely to be proportional to the beta functions of the couplings. With the detailed steps illustrated in 
sec.  IV of the supplemental materials \cite{supp}, 
the conformal Ward-Takahashi identities on the projected $2$-dimensional space $(x_\perp =0)$ can be written as
 \begin{widetext}
\begin{align}\label{eq:WI1}
\sum_{i=1}^n &\left[ y_{i} \cdot  \der_{y_i} +\Delta_\phi\right] G_n(y_1,\cdots,y_n) \no\\
&=  \frac{\beta_h}{4!} \mu^{2\vareps} \int d^2y\, G_n(y_1,\cdots,y_n; [\phi^4]) + \frac{\beta_g}{2} \mu^{\vareps} \int d^2y\, \delta(y_\perp) G_n(y_1,\cdots,y_n; [\hat \phi^2])\,,\\
\label{eq:WI2}
\sum_{i=1}^n &\left[ (2y_{i}^\mu y_i^a - \delta^{\mu a} y_i^2) \frac{\der}{ \der y_i^\mu} +2\Delta_\phi y_i^a\right] G_n(y_1,\cdots,y_n)\no\\
&=2 \frac{\beta_h}{4!} \mu^{2\vareps} \int d^2y \, y^a  \, G_n(y_1,\cdots,y_n; [\phi^4]) + 2\frac{\beta_g}{2} \mu^{\vareps} \int d^2y \,y^a \, \delta(y_\perp) G_n(y_1,\cdots,y_n; [\hat \phi^2])\,,
\end{align}
\end{widetext}
where $G_n(y_1,\cdots,y_n)$ is the renormalized $n$-point function of the LG field $[\phi]$. The $\mu,\nu$, and $a$ indices run over the $2$-dimensional bulk and the interface respectively. The first equality is related to the scaling transformation, while the second one shows the response to the special conformal transformation along the interface. At the IR fixed point, the vanishing beta functions restore the scale and special conformal Ward-Takahashi identities, hence proving the conformal invariance.\footnote{One might worry about a potential ``$\infty$'' coming from the integral part, such that $0\times \infty \sim O(1)$. In \cite{Paulos:2015jfa}, it was argued that a cancellation should happen between the UV and IR parts to ensure that the integral times the scale is actually of $O(1)$.}

Now we pay special attention to the displacement operator of the interface \cite{McAvity:1993ue,McAvity:1995zd,Herzog:2017xha}. Though a locally conserved stress tensor is absent in the LRI model, the displacement operator can still be constructed from the higher-dimensional theory.
It appears in the divergence  of the stress tensor as follows
\begin{align}\label{eq:dTMN}
 \der^M T_{MN} = - E_N &+ \delta^{(D-d)}(x_\perp) D_N \no\\
& + \delta^{(D-r)}(x_\perp;y_\perp) \bar D_N\,,
\end{align}
with $N$ running over the full $D$-dimensional space. The first term $E_N$ vanishes on-shell, and the two delta-localized terms give rise to two displacement operators $D_{N}$ and $\bar D_N$, of which $\bar D_N$ is the relevant one for the LRI interface, while $D_N$ displaces the full interface LRI CFT inside the ambient space.  All three are renormalized operators whose explicit expressions are given as
\begin{align}\label{eq:EN}
E_N &= \der_N \Phi \left( -\Box \Phi  +  \delta^{(D-d)} (x_\perp) \frac{h}{3!} [\phi^3] 
\right.\no\\
&~~~~~~~~~~~~~~\left.+  \delta^{(D-r)} (x_\perp;y_\perp)~{g} [\hat\phi] \right)\,,\\
D_{N} &= \left(\frac{h}{4!}  [\der_N \phi^4] + 
\delta^{(d-r)}(y_\perp)\frac{g}{2} [\der_N \hat \phi^2] \right)\delta_{N,n_1} 
\,,\\
\label{eq:LRIdisplacement}
\bar D_{N} &= \frac{g}{2}  [\der_N \hat \phi^2] \, \delta_{N,n_2}\,.
\end{align}
The terms inside the bracket of \eqref{eq:EN} are e.o.m. related, $n_1$  denotes the normal direction towards the ambient space while $n_2$ denotes the normal direction to the LRI interface. The two displacement operators are protected with exact dimensions, 
\be
\Delta_{D_N} = d+1= 3\,,~~~ \Delta_{\bar D_N}  = r+1=2\,,
\ee
which agree with the perturbative checks through their anomalous dimensions.
 Considering the renormalization factor of $\der_{n_2} \hat\phi^2 = Z_{\bar D_N} [\der_{n_2} \hat\phi^2]$, at the linear order in the couplings 
\be
Z_{\bar D_N} = 1- \frac{g}{2 \pi \vareps} - \frac{h}{32\pi \vareps}\,~~~\Rightarrow ~~~\gamma_{\bar D_N}^* =\vareps\,,
\ee
consistent with its protected dimension. Renormalizing $\der_{n_1} \phi^4$ involves a mixture of the renormalized operators $[\der_{n_1} \phi^4]$ and $[\der_{n_1} \hat\phi^2]$
\begin{align}
\begin{pmatrix}
\frac{1}{4!} \der_{n_1} \phi^4 \\
\delta(y_\perp) \frac{1}{2} \der_{n_1} \hat\phi^2
\end{pmatrix}
 = Z_{\text{mix}}
\begin{pmatrix}
\frac{1}{4!} [\der_{n_1} \phi^4] \\
\delta(y_\perp) \frac{1}{2} [\der_{n_1} \hat\phi^2]
\end{pmatrix} \,,
\end{align}
with the renormalization factor matrix given as
\be
Z_{\text{mix}}=
\begin{pmatrix}
1- \frac{3h}{32\pi \vareps} & - \frac{g}{32\pi\vareps}\\
0 & 1- \frac{g}{2\pi \vareps}
\end{pmatrix}\,.
\ee
The off-diagonal entry cancels exactly the mixed coupling term from the renormalization factor in $g_0$ at this order, such that $D_N$ remains finite even when written in terms of the bare fields and couplings. The eigenvalues of the anomalous dimension matrix $\gamma_{\text{mix}} = \der \ln Z_{\text{mix}}/\der\ln \mu$ give the anomalous dimension of $D_N$ and
$[\der_{n_1}\hat\phi^2]$, $2\vareps$ and $\vareps/3$ respectively, in the leading order in $\vareps$. This leading correction is consistent with the protected dimension of $D_N$ as $\left( \frac{D}{2}-1\right)\times 4 +1 +2\vareps = 3 = \Delta_{D_N}$.

Finally, we would like to study the (non-)factorizing property of our interface in the IR limit. The residual conformal symmetries fix the bulk two-point function of scalar operators on the two sides of the interface to be $\langle O(y_1) O(y_2) \rangle= \frac{1}{|y_{1,\perp}|^{\Delta}|y_{2,\perp}|^\Delta} G(\xi)$, where $G(\xi)$ is a function of the cross-ratio $\xi = \frac{  (y_{1}-y_{2})^2}{4 |y_{1,\perp}| |y_{2,\perp}|}$ \cite{Billo:2016cpy,Gadde:2016fbj,Dey:2020jlc,Okuyama:2023fge,Shimamori:2024yms} and factorization requires $G(\xi) = \text{const}$. However,
it is hardly satisfied for any bulk operators in the LRI theory.

Taking $\mO (y) =  [\phi^n](y)$, the perturbative calculations show $G(\xi) = \frac{C_{\mO\mO}}{\xi^{\Delta}} + \mathcal{O}(h,g) $ as in 
sec. II of the supplemental materials \cite{supp}, where the $h$-term  shifts the dimension $\Delta$ and the $g$-term in the $\xi \to 0$ limit is related to the leading correction to the bulk-to-defect coefficient $\mB_{\phi^n \hat\phi^n}$ as in \eqref{eq:Btdcoeff}. 
Obviously the factorization condition cannot be satisfied perturbatively.
Thus our interface seems not to comply with the factorization proposal. The technical reason why the proof in \cite{Popov:2025cha} does not apply here is that the relevant deformation $g\phi^2$ has a non-trivial fixed point at a finite value of $g$; 
taking $g\to \infty$ is not guaranteed
as what has been assumed in their proof. More physically, the space is not factorized because the two halves can be still connected via the ``extra dimensions".
 
Let us comment on the relation between the displacement operator and the factorization property. The explicit computation shows that the two-point function coefficient of the displacement operator is perturbatively small of $O(\vareps^2)$, which should be in accordance with the non-factorization property of our interface.
We suspect that the Zamolodchikov norm of the displacement operator (called $C_D$ in the literature) may serve as a diagnostic of the factorization property of the interface. One can expect once this coefficient saturates its upper bound, the translational symmetry would be maximally broken, hence the factorization. It is the case in two-dimensional interface CFTs with local Virasoro symmetries, where factorization indicates that $C_D = 2( c_L +c_R)$ \cite{Quella:2006de,Meineri:2019ycm}, saturating its upper bound. However, how this criterion can apply to a non-local and/or higher-dimensional interface CFT requires further exploration.


\acknowledgements
 
We would like to thank Shota Komatsu, Sungjay Lee, Marco Meineri, Tatsuma Nishioka, Miguel Paulos, Massimo Porrati, Slava Rychkov, Satoshi Yamaguchi, Piljin Yi and Tadashi Takayanagi for interesting discussions and inspirations at various stages.  We would like to thank the anonymous referee for suggesting the simple quantum mechanical model considered in sec. III of the supplemental materials. We also  thank Kyushu University Institute for Advanced Study and RIKEN Interdisciplinary Theoretical and Mathematical Sciences Program. Discussions during the ``Kyushu IAS-iTHEMS workshop: Non-perturbative methods in QFT" were useful in completing this work. DG wants to thank Yukawa Institute for  Theoretical Physics (YITP) for hospitality during his regular visits, as well as Korea Institute for Advanced Study (KIAS) during the final stage of this work, where part of the results were presented.
DG is supported in part by the JSPS Grant-in-Aid for Transformative Research Areas (A) “Extreme Universe” No. 21H05182 and No. 21H05190. 
YN is supported in part by the JSPS KAKENHI Grant No. 21K03581.



\bibliography{bibLRI}


 \onecolumngrid
 \appendix

\section{Correlator in the $r$-dimensional space}
We can evaluate the bulk-bulk correlator in the free theory and perform a Fourier transformation to the parallel $r$-dimensional momentum space,
\begin{align}
&~~\langle \Phi(\bar p_1,x_{ 1_\perp}) \Phi(\bar p_2,x_{ 2_\perp}) \rangle \no\\
&= C_{\Phi\Phi} \int d^r z_1\, d^rz_2 \frac{e^{i\bar p_1 z_1 + i \bar p_2 z_2}}{\left( z_{12}^2 + x_{12_\perp}^2\right)^{\Delta_\Phi}}\no\\
&= \frac{1}{4\pi^{D/2}} \int _0^{\infty} ds s^{\frac{D}{2}-2}e^{-s x_{12_\perp}^2} \int d^r z_1 d^r z_2 e^{-s z_{12}^2 + i\bar p_1 z_1 + i\bar p_2 z_2}\no\\
&= \frac{(2\pi)^r \delta ^{(r)}(\bar p_1+\bar p_2)}{4\pi ^{\frac{D-r}{2}}} \int_0^\infty  ds s^{\frac{D-r}{2} -2} e^{-s x_{12_\perp}^2 - \frac{\bar p_1^2}{4s}} \no\\
&= \frac{(2\pi)^r \delta ^{(r)}(\bar p_1+\bar p_2)}{(2\pi) ^{\frac{D-r}{2}}}  \left( \frac{|\bar p_1|}{|x_{12_\perp}|}\right)^{\frac{D-r-2}{2}} K_{\frac{2+r-D}{2}}(|\bar p_1| |x_{12_\perp}|)\,,
\end{align}
where $C_{\Phi\Phi} = \frac{\Gamma((D-2)/2)}{4\pi^{D/2}}$ is the two-point coefficient for the free field $\Phi(x)$ and $K_\alpha (x)$ is the modified Bessel function of the second kind.\footnote{This correlator resembles the Green function for the massless graviton mode in the transverse space to the brane world discussed in \cite{Dvali:2000xg}. We thank {\it Massimo Porrati} for mentioning this paper.}
In the above derivation, we have used  the  following Laplace transformation 
\be
\frac{1}{y^\Delta} = \frac{1}{\Gamma(\Delta)} \int_0^{\infty} e^{-s y } s^{\Delta -1}ds\,,
\ee
and the integral 
\be
 \int d^r z e^{-s z^2} = \left(\frac{\pi}{s}\right)^{r/2}\,.
\ee
In the limit $x_{1_\perp} =x_{2_\perp} = 0$, we obtain the correlator in the parallel momentum space for $D-r<2$\footnote{We thank {\it Satoshi Yamaguchi} for discussion on this point.}
\be\label{eq:parallelS}
\langle \Phi(\bar p_1,0) \Phi(\bar p_2,0) \rangle = (2\pi)^r \delta ^{(r)}(\bar p_1+ \bar p_2)\frac{a_{D,r}}{\bar p_1^{2+r-D}}\,, \ee
as
$
K_{\alpha}(x) \stackrel{x\to 0}{\sim} \frac{1}{2} \Gamma(|\alpha|) (2/x)^{|\alpha|}
$.


\section{Bulk two-point function}\label{app:Bulk2pt}
Let us start by considering the  Feynman reparametrization for the following integral
\begin{align}
I_{r,\alpha,\beta}(y_1,y_2)&=\int d^rz \frac{1}{|y_1-z|^{2\alpha} |y_2 -z|^{2\beta}} \no\\
&= \int d^rz \frac{1}{(y_{1_\perp}^2+(z-z_{12})^2)^{\alpha} (y_{2_\perp}^2 + z^2)^{\beta}}\no\\
&= \frac{\Gamma(\alpha+\beta)}{\Gamma(\alpha)\Gamma(\beta)}\int_0^1 du \int d^rz
\frac{u^{\alpha-1}(1-u)^{\beta-1}}{\left((z- u z_{12})^2 +u(1-u)z_{12}^2 +(1-u) y_{1_\perp}^2 + u y_{2_\perp}^2 \right)^{\alpha+\beta}}\no\\
&= \frac{\pi^{r/2}\Gamma(\alpha+\beta-r/2)}{\Gamma(\alpha)\Gamma(\beta)}
\int_0^1 du \frac{u^{\alpha-1}(1-u)^{\beta-1}}{\left( u(1-u)z_{12}^2 +(1-u) y_{1_\perp}^2 + u y_{2_\perp}^2 \right)^{\alpha+\beta-r/2}} \ . 
\end{align}
Since we are mainly concerned with the transverse distance, the above integral can be simplified by taking  $z_{12} = 0$, which gives
\be
I_{r,\alpha,\beta}(y_{1\perp},y_{2\perp})=
\frac{\pi^{r/2}\Gamma(\alpha+\beta -r/2)}{\Gamma(\alpha+\beta)}\frac{\,
   _2F_1\left(\alpha,\alpha+\beta-r/2;\alpha+\beta ; 1 - \frac{y_{2_\perp}^2}{y_{1_\perp}^2} \right)}{y_{1_\perp}^{2(\alpha+\beta)-r}}\,.
\ee

For the bulk two-point function up to first order in the couplings
\begin{align}
\langle \phi^n (y_1) \phi^n(y_2)\rangle = 
\frac{n! C_{\Phi\Phi}^n}{|y_{12}|^{2n\Delta_\Phi}}& 
- h_0 \frac{n(n-1)\,n!}{4|y_{12}|^{2(n-2)\Delta_\Phi}}
 \int d^d{  y} \frac{C_{\Phi\Phi}^{n+2}}{|  y -   y_1|^{4\Delta_\Phi} |  y -   y_2|^{4\Delta_\Phi}} \no\\
& -g_0 \frac{n \,n!}{|y_{12}|^{2(n-1)\Delta_\Phi}}
 \int d^r{  z} \frac{C_{\Phi\Phi}^{n+1}}{|  z -   y_1|^{2\Delta_\Phi} |  z -   y_2|^{2\Delta_\Phi}} \ .
\end{align}

Considering the renormalization factor of $\phi^n$
\be
Z_{\phi^n} = 1 - \frac{ (n-1) n \Gamma \left(\frac{d}{4}\right)^2}{64 \pi ^2 \Gamma
   \left(\frac{d}{2}\right)} \frac{h}{\vareps} +\dots \ ,
\ee
we obtain the renormalized two-point function in $d=2$ for $z_{12}=0$
\begin{align}
\langle [\phi^n] (y_1) [\phi^n](y_2)\rangle =\frac{n!}{\left( 4\pi \right)^{n}y_{12_\perp}^{n}}  &\left[ 1+  n \varepsilon \left(
   \ln (y_{12_\perp}) +\frac{\gamma +\ln(4\pi)}{2} \right) \right. 
\no\\
- &h_* \frac{n(n-1)}{16\pi}
   \left(
   \ln (\mu\, y_{12_\perp}) +\frac{\gamma +\ln(4\pi)}{2} \right) \no\\
   -&\left.
   g_*\frac{ n}{2\pi}\left|1- \frac{y_{2_\perp}}{y_{1_\perp}} \right|
   K\left(1-\frac{y_{2_\perp}^2}{y_{1_\perp}^2}\right) 
   \right] \ ,
\end{align}
where $K(x) = \frac{\pi}{2} \, _2F_1(1/2,1/2;1;x)$ is the complete elliptic integral of the first kind.

\section{A simple quantum mechanical model}
We consider a quantum mechanical model in $1+1$ dimension with a fractional kinetic term and a delta-type potential, which has been considered in \cite{10.1063/1.2749172,10.1063/1.3525976}. The Hamiltonian has the form
\be
H = D_\alpha (-\der_x^2)^{\alpha/2} + V_0\, \delta(x)
\ee
where $D_\alpha$ is the generalized diffusive constant with mass dimension $[m]^{1-\alpha}$ $(1<\alpha\le 2)$ and $(-\der_x^2)^{\alpha/2}$ is the fractional Riesz derivative. 
The fractional Schrödinger equation then reads (taking $\hbar =1$)
\be
i \der_t \psi =  D_\alpha (-\der_x^2)^{\alpha/2} \psi + V_0 \delta(x)  \psi\,,
\ee
or in the momentum space
\be
\left( |p|^\alpha - \lambda^\alpha \right)\, \phi(p) = - \frac{V_0}{D_\alpha} \psi(0)
\ee
with $\lambda^\alpha \equiv \frac{\omega}{D_\alpha}$ and $\psi(x) \equiv \int \frac{dp}{2\pi} \phi(p) e^{ipx} $. Assuming $\lambda \ge 0$ and using the Lippmann-Schwinger formalism to select the outgoing scattering solution, we have
\be\label{eq:psimomentumspace}
\phi( p) =  2\pi \delta (p-\lambda)- \frac{V_0}{D_\alpha} \frac{ \psi(0)}{|p|^\alpha - \lambda^\alpha -i \eps} \,,
\ee
where the $-i\eps$ choice ensures the correct boundary condition. Transforming back to the position space gives
\be
\psi(x) = e^{i\lambda x} - \frac{V_0}{D_\alpha} \psi(0) G(x,\omega)\,,~~~ G(x,\omega) \equiv \int\frac{dp}{2\pi} \frac{e^{i p x}}{|p|^\alpha - \lambda^\alpha - i\eps} \,.
\ee
To obtain $\psi(0)$, we need to get $G(0,\omega)$, which can be done by 
separating the integral into a real part and an imaginary part using the Plemelj formula, 
\be
 G(0,\omega) =\int \frac{d p}{2\pi} \frac{1}{|p|^\alpha - \lambda^\alpha - i\eps} = \frac{1}{\alpha  \lambda^{\alpha-1}} \left(-\cot \frac{\pi}{\alpha} +i \right)
  \,,
\ee
substuting back to the solution for the wavefunction $\psi(x)$ gives
\be
\psi(0) = \frac{1}{1+\frac{V_0}{D_\alpha} G (0,\omega)}\,.
\ee
We are primarily interested in the reflection coefficient in this scattering problem, for that, we need to obtain the asymptotic behavior of the wavefunction at $x\to -\infty$. For $x<0$, $G(x,\omega)$ can be evaluated by closing the contour in the lower half plane clockwisely,
\be
G(x,\omega) = - \frac{i}{\alpha \lambda^{\alpha-1}}e^{ i\lambda x}\,,~~~\text{for}~~~x<0
\ee
then the reflective coefficient can be evaluated as
\be
R = -i \frac{V_0}{D_\alpha} \frac{1}{\alpha \lambda^{\alpha-1} + \frac{V_0}{D_\alpha}(- \cot \frac{\pi}{\alpha} + i) }\,.
\ee
For the lower energy modes $\omega \to 0$ or $\lambda \to 0$, the reflective probability shows
\be
|R|^2 = \sin ^2 \left(\frac{\pi}{\alpha} \right) <1\,,~~~\text{for}~~~1<\alpha<2
\ee
which indicates that due to the super-diffusive nature of the Lévy flight (non-locality), particles have a finite probability of tunneling through (or hopping over) a delta potential barrier even with zero incident energy. This is in sharp contrast to standard quantum mechanics $(\alpha = 2)$,
where total reflection occurs.

The above quantum mechanical model with a fractional  kinetic term illustrates the essence of the non-factorizing interface in the main text. 


\section{Derivation of the  Ward-Takahashi identities}\label{sec:WardIdRG}
\subsection{A mixture of operators}
For the $n$-point function built by $\Phi$, there is no distinction between the renormalized ones and the bare ones as the equivalent higher-dimensional theory (2) is free, therefore
\be
G_{n}^{(0)} (x_1,x_2,\dots, x_n) =\langle \Phi (x_1) \Phi(x_2) \cdots \Phi(x_n) \rangle
= G_{n}(x_1,x_2,\dots, x_n)\,.
\ee
The bare couplings can be written  as in  \cite{Brown:1979pq} 
\begin{align}
h_0 = \mu^{2\vareps} h \exp (\tilde f (h))\,,~~~  g_0 =  \mu^{\vareps - \frac{d}{2}+r} g \exp(f (h,g))\,,
\end{align}
with both functions $f(h,g)$ and $\tilde f (h)$ being a sum of poles in $\vareps$, explicitly  as
\be
\tilde f (h)= \sum_{i=1} \frac{\tilde f_i(h)}{\vareps^i}\,,~~~ f (h,g) = \sum_{i=1} \frac{f_i (h,g)}{\vareps^i}\,.
\ee
As the bare couplings are independent of the renormalization scale $\mu$, we have the following two equations 
\begin{align}
\mu \frac{\der h_0}{\der\mu} =0 ~~~&\Rightarrow ~~~ 2\vareps h + \beta_h (1+ h\, \der_h \tilde f) =0\,,\\
\mu \frac{\der g_0}{\der\mu} =0 ~~~&\Rightarrow ~~~ \left( \vareps -\frac{d}{2} +r \right)g +\beta_g +g (\beta_h\, \der_h f + \beta_g\, \der_g f)=0\,.
\end{align}
Requiring the beta functions to be free of $\vareps$ poles relates $\tilde f_k$ to its lower order counterparts, namely $\{\tilde f_1,\dots,\tilde f_{k-1}\}$, and $f_k$ to a mixture of $\{\tilde f_1,\dots,\tilde f_{k-1}\}$ and $\{f_1,\dots, f_{k-1} \}$, up to integration constants. In this manner, those two beta functions are entirely determined by $f_1$ and $\tilde f_1$,
\begin{align}
    \beta_h &= - 2\vareps h + 2h^2 \der_h \tilde f_1\,,\\
    \beta_g &= -\left( \vareps -\frac{d}{2} +r \right) g +g^2 \der_g f_1 + 2gh\der_h f_1\,.
\end{align}
With these two equations, the partial derivatives can be worked out, given as
\begin{align}
\frac{\der g_0}{\der g} &= g_0 \left( \frac{1}{g} + \frac{\der f}{\der g}\right)\,,~~~
 \frac{\der g_0}{\der h} = g_0 \frac{\der f}{\der h}\,,\\
 \frac{\der h_0}{\der h} &= - \frac{2\vareps h_0}{\beta_h}\,.
\end{align}
By construction, the renormalized correlator $G_{n}(x_1,x_2,\dots, x_n)$ is finite as $\vareps \to 0$, but we can also require that their derivatives w.r.t. the renormalized couplings should also be finite, which offers a way to renormalize the composite operators $h_0 \phi^4$ and $g_0 \hat\phi^2$  (taking $r=d/2$)
\begin{align}
&\int d^rx\, G_n \left(x_1,\dots, x_n; - \frac{\mu^{\vareps}}{2}[\hat \phi^2] \right) \stackrel{!}{=}\frac{\der}{\der g} G_n(x_1,\dots, x_n)
\no\\
& = \frac{\der g_0}{\der g} \frac{\der }{\der g_0} G_{n}^{(0)} (x_1,\dots, x_n)  
\stackrel{!}{=} \frac{\der g_0}{\der g} \int d^rx\, G_n ^{(0)}\left(x_1,\dots, x_n; - \frac{1}{2}\hat \phi^2 \right)\,,\\
&\int d^dx\,G_n \left(x_1,\dots, x_n; - \frac{\mu^{2\vareps}}{4!}[\phi^4] \right) \stackrel{!}{=}\frac{\der}{\der h} G_n(x_1,\dots, x_n)= \left( \frac{\der g_0}{\der h} \frac{\der }{\der g_0} +  \frac{\der h_0}{\der h} \frac{\der }{\der h_0} \right)\, G_{n}^{(0)} (x_1,\dots, x_n) 
\no\\
&  
\stackrel{!}{=} \frac{\der g_0}{\der h} \int d^d x\, G_n ^{(0)}\left(x_1,\dots, x_n; - \frac{1}{2}\hat \phi^2 \delta^{(d-r)}(y_\perp) \right) 
+ \frac{\der h_0}{\der h}  \int d^d x\,G_n ^{(0)}\left(x_1,\dots, x_n; - \frac{1}{4!} \phi^4  \right)\,.
\end{align}
Substituting the explicit expressions of the derivatives into the  above two equations gives the relations between the renormalized composite operators and the bare ones
\begin{align}\label{eq:RGphi2}
\frac{1}{2} g_0\hat \phi^2  &= \left(  \frac{1}{g} + \frac{\der f}{\der g}\right)^{-1}\frac{\mu^\vareps}{2}[\hat \phi^2] \,,\\
\label{eq:RGphi4}
\frac{1}{4!}h_0 \phi^4 &= -\frac{\beta_h}{2\vareps}  \frac{\mu^{2\vareps}}{4!}[\phi^4] +
 \frac{\beta_h}{2\vareps} \frac{\der f}{\der h} \delta^{(d-r)}(y_\perp) \frac{1}{2} g_0\hat \phi^2\,.
\end{align}
Thus one sees that renormalizing the operator $h_0 \phi^4$ involves a mixture of the renormalized operators $[\phi^4]$ and $[\hat \phi^2]$, while the renormalization of the operator $g_0 \hat \phi^2$ is solely related to $[\hat \phi^2]$. 
More explicitly in our case $r=\frac{d}{2}=1$, 
\be
\begin{pmatrix}
    \frac{1}{4!}h_0 \phi^4 \\[1ex]
    \frac{1}{2} g_0\hat \phi^2 \delta(y_\perp)
\end{pmatrix} 
=\bar Z_{\text{mix}} 
\begin{pmatrix}
    \frac{\mu^{2\vareps}}{4!}h [\phi^4] \\[1ex]
    \frac{\mu^\vareps}{2} g [\hat \phi^2] \delta(y_\perp)
\end{pmatrix} 
\,,~~~
\text{or}
~~~
\begin{pmatrix}
    \frac{1}{4!} \phi^4 \\[1ex]
    \frac{1}{2}\hat \phi^2 \delta(y_\perp)
\end{pmatrix} 
=Z_{\text{mix}} 
\begin{pmatrix}
    \frac{1}{4!} [\phi^4] \\[1ex]
    \frac{1}{2} [\hat \phi^2] \delta(y_\perp)
\end{pmatrix} 
\ee
where these two renormalization matrices $\bar Z_{\text{mix}}$ and $Z_{\text{mix}}$ are related as
\be
Z_{\text{mix}} = 
\begin{pmatrix}
    h_0^{-1} &0\\
    0& g_0^{-1}
\end{pmatrix}\bar Z_{\text{mix}}
\begin{pmatrix}
    \mu^{2\vareps}h &0\\
    0& \mu^\vareps g
\end{pmatrix}\,.
\ee
Up to order $O(\vareps^{-1})$, the renormalization matrix $Z_{\text{mix}}$ is found to be 
\be
Z_{\text{mix}} = 
\begin{pmatrix}
    1 -  \frac{1}{\vareps}\frac{3h}{16\pi} & - \frac{1}{\vareps}\frac{g}{32\pi}\\[1ex]
    0 & 1- \frac{1}{\vareps}\frac{h+32g }{32\pi} 
\end{pmatrix}+ O(g^2,h^2,gh)\,,
\ee
giving the anomalous dimension matrix at linear order in the couplings
\be
\gamma_{\text{mix}} = Z_{\text{mix}}^{-1}\, \mu \der_\mu Z_{\text{mix}}  = 
\begin{pmatrix}
 \frac{3 h}{8 \pi } & \frac{g}{32 \pi } \\[1ex]
 0 & \frac{h+16g}{16 \pi } 
\end{pmatrix}+ \text{higher order terms}\,,
\ee
where its eigenvalues are the anomalous dimensions of two scaling operators, which are consistent with equations (7) and (21) in the main text.

\subsection{Broken Ward-Takahashi identities along the RG} 
To derive the  Ward-Takahashi identities along the RG flow, we can consider the stress  tensor in the higher-dimensional theory,
\begin{align}
T_{MN} &= \der_M \Phi \der_N \Phi - \frac{1}{2} \delta_{MN}(\der_K \Phi)^2 
-\delta_{MN}^{\parallel_d} \delta^{(D-d)} (x_\perp)\frac{h_0}{4!} \Phi^4
 -\delta_{MN}^{\parallel_r} \delta^{(D-r)} (x_\perp;y_\perp)\frac{g_0}{2!} \Phi^2\,,
\end{align}
where the capital indices $M,N...$ runs  in all the $D$-dimensional space, $\delta_{MN}^{\parallel_d}$ and  $\delta_{MN}^{\parallel_r}$  project the indices to the parallel directions of the hyperplanes $x_\perp =0$ and $ x_\perp = y_\perp =0$ respectively. Its trace is given by
\be
{T^M}_{M} = - \left( \frac{D}{2} -1\right) E + (2D-4-d)  \delta^{(D-d)} (x_\perp)\frac{h_0}{4!} \Phi^4 + (D-2-r) \delta^{(D-r)} (x_\perp;y_\perp)\frac{g_0}{2!} \Phi^2 - \frac{D-2}{4}\der_K^2 \Phi^2\,,
\ee
where the first term is related to the e.o.m. of $\Phi$ and is given as
\begin{align}
E &= \Phi \left( -\der_K^2 \Phi +  \delta^{(D-d)} (x_\perp) \frac{h_0}{3!} \Phi^3 
+  \delta^{(D-r)} (x_\perp;y_\perp){g_0} \Phi \right)\,,
\end{align}
where terms inside the bracket constitute the e.o.m. for $\Phi$. Consider the renormalizations \eqref{eq:RGphi2} and \eqref{eq:RGphi4}, the trace of the stress tensor has the form
\be
{T^M}_{M} = -\Delta_\phi E + \frac{\beta_h}{4!} \mu^{2\vareps} \delta^{D-d}(x_\perp) [\phi^4] + \frac{\beta_g}{2} \mu^{\vareps} \delta^{D-r}(x_\perp;y_\perp) [\hat \phi^2] -  \frac{D-2}{4}\der_K^2 \Phi^2\,.
\ee
Magically, the terms related to $[\phi^2]$ coming from the bare operators $h_0 \phi^4$ and $g_0 \hat \phi^2$ get recombined to a single term proportional to the beta function of $g$. 
 The dilatation and special conformal currents can be constructed in terms of the stress tensor as 
\be
\mD_M = T_{MN}\,x^N\,,~~~ {\mC_{M}}^L = T_{MN}(2x^N x^L - \delta^{NL}x^2)\,,
\ee
their divergences can be calculated using the divergence of the stress tensor (30),
\begin{align}\label{eq:dDilaton}
\der^{M} \mD_{M} &= - x^{N}\,E_N + {T^M}_{M}\,,\\
\label{eq:dSpecial}
\der^{M} {\mC_{M}}^L &=  -(2x^N x^L - \delta^{NL}x^2)(E_N - \delta^{(D-d)}(x_\perp) D_N - \delta^{(D-r)}(x_\perp;y_\perp) \bar D_N) +2x^L\,  {T^M}_{M}\,,
\end{align}
where the displacements $D_N$ and $\bar D_N$ measure the brokenness of the special conformal currents in the direction transverse to the $r$-dimensional defect. However, if we focus on the directions parallel to the $r$-dimensional defect (taking $L = a$), they become irrelevant.
Inserting \eqref{eq:dDilaton} and \eqref{eq:dSpecial} into the $n$-point function $G_n$ and integrating over the full $D$-dimensional space of the equivalent higher-dimensional theory, we obtain 
\begin{align}
\sum_{i=1}^n &\left[ x_{i} \cdot  \der_{x_i} +\Delta_\phi\right] G_n(x_1,\cdots,x_n) \no\\
&=  \frac{\beta_h}{4!} \mu^{2\vareps} \int d^Dx\, \delta^{(D-d)}(x_\perp)G_n(x_1,\cdots,x_n; [\phi^4]) + \frac{\beta_g}{2} \mu^{\vareps} \int d^Dx\, \delta^{(D-r)}(x_\perp;y_\perp) G_n(x_1,\cdots,x_n; [\hat \phi^2])\,,\\
\sum_{i=1}^n &\left[ (2x_{i}^M x_i^a - \delta^{Ma} x_i^2) \frac{\der}{ \der x_i^M} +2\Delta_\phi x_i^a\right] G_n(x_1,\cdots,x_n)\no\\
&=2 \frac{\beta_h}{4!} \mu^{2\vareps} \int d^Dx \, x^a\delta^{(D-d)}(x_\perp)  \, G_n(x_1,\cdots,x_n; [\phi^4]) + 2\frac{\beta_g}{2} \mu^{\vareps} \int d^Dx \,x^a \, \delta^{(D-r)}(x_\perp;y_\perp) G_n(x_1,\cdots,x_n; [\hat \phi^2])\,.
\end{align}
Upon taking the limit ${x_i}_\perp \to 0$, which is a safe limit as the fundamental field $\phi$ 
does not acquire an anomalous dimension, the above two broken Ward-Takahashi identities give rise to (28) and (29) respectively.



\end{document}